\documentclass[structabstract]{aa}  
\usepackage{natbib}
\bibpunct{(}{)}{;}{a}{}{,} 

\usepackage{graphicx}
\usepackage{natbib}
\usepackage{color}
\usepackage{rotating}
\usepackage{txfonts}
\begin{document}
\definecolor{grey}{rgb}{0.5,0.6,0.7}
\definecolor{darkred}{rgb}{0.5,0.0,0.0}
\definecolor{darkgreen}{rgb}{0.0,0.7,0.0}
\newcommand{\euge}[1]{{\bf\textcolor{darkred}{Euge says: #1}}}
\newcommand{\claudia}[1]{{\bf\textcolor{blue}{Claudia says: #1}}}
\newcommand{\hernan}[1]{{\bf\textcolor{darkgreen}{Hern\'an says: #1}}}
\newcommand{\hide}[1]{\textcolor{grey}{#1}}%

   \title{Fossil Groups in the Millennium Simulation}

   \subtitle{ Evolution of the Brightest Galaxies}

   \author{Eugenia D\'{\i}az-Gim\'enez\inst{1},
	   Hern\'an Muriel\inst{1}	
          \and
          Claudia Mendes de Oliveira\inst{2}
          }

   \institute{IATE (CONICET-UNC) \& OAC (UNC). Laprida 854, C\'ordoba 5000. Argentina\\
              \email{eugeniadiazz@gmail.com, hernan@oac.uncor.edu}
         \and
             IAG, USP. Rua do Mat\~ao 1226, S\~ao Paulo. Brazil\\
             \email{oliveira@astro.iag.usp.br}
             }

   \date{Received March 11, 2008; accepted August 19, 2008 }

 
  \abstract
   {
   }
   { 
We create a catalogue of \emph{simulated} fossil groups and study their properties, in particular the merging histories of their first-ranked galaxies. We compare the \emph{simulated} fossil group properties with those of both \emph{simulated} non-fossil and \emph{observed} fossil groups.
   }
   {Using simulations and a mock galaxy catalogue, we searched for massive ($>$ 5 $\times$ 10$^{13} \ h^{-1} {\cal M}_\odot$) fossil groups in
the Millennium Simulation Galaxy Catalogue.
In addition, attempted to identify observed fossil groups in the Sloan Digital Sky 
Survey Data Release 6 using identical selection criteria.
   }
   { Our predictions on the basis of the simulation data are:
(a) fossil groups comprise about 5.5\% of the total population of 
groups/clusters with masses larger than 
5 x 10$^{13} \ h^{-1} {\cal M}_\odot$. This fraction is consistent with
the fraction of fossil groups identified in the SDSS, after all observational 
biases have been taken into account;
(b) about 88\%
of the dominant central objects in fossil groups
are elliptical galaxies that have a median R-band absolute magnitude of
$\sim -23.5-5 \ log \ h$, which is typical of the observed fossil groups known 
in the literature; 
(c) 
first-ranked galaxies of systems with
$ {\cal M} >$   5 x 10$^{13} \ h^{-1} {\cal M}_\odot$, 
regardless of whether they are either fossil or
non-fossil,
are mainly formed by gas-poor mergers; 
(d) although
fossil groups, in general, assembled most of their virial masses at
higher redshifts in comparison with non-fossil groups, 
\emph{first-ranked 
galaxies in fossil groups merged later, 
i.e. at lower redshifts}, 
compared with
their non-fossil-group counterparts.
   }
   {We therefore expect to observe a number of luminous
galaxies in the centres of fossil groups that show signs of a recent
major merger.}


   \authorrunning{D\'{\i}az-Gim\'enez et al.}
   \maketitle
%

\section{Introduction}
\cite{Jones03} identified fossil groups as spatially extended X-ray sources 
with an X-ray luminosity $L_X>10^{42} \ h_{50}^{-2} \ erg \ s^{-1}$ whose 
optical counterpart was a bound system of galaxies with $\Delta M_{12}>2$ mag, where
$\Delta M_{12}$ was the difference in absolute magnitude in R-band between the brightest 
and the second brightest galaxies in the system within half the projected virial 
radius ($r_{vir}$). The dynamical masses of the systems studied so far 
are comparable to those of 
rich clusters ($\sim 10^{13}-10^{14} \ h^{-1} {\cal M}_\odot$) \citep{Mendes06,cyp06,Mendes08, KPJ06}. 
Fossil groups may be of considerable importance as the place of formation of a significant
fraction of all giant ellipticals.
Beside minor differences in the definition of fossil groups, 
their incidence rate was estimated 
by observational, analytical, numerical, and semi-analytical analyses.
\cite{Vik99} and \cite{Jones03} stated that fossil groups represent (8-20)\% of observed systems in 
the same mass range. \cite{vandenbosch07} used the 2dF Galaxy Redshift Survey data to measure
a fossil fraction of 6.5\% among groups with masses $(10^{13}-10^{14}) \ h^{-1} {\cal M}_\odot$.
\cite{M06} estimated analytically that fossil groups represent 5-40\%
of groups with masses in the range $\sim 10^{13}-10^{14} \ h^{-1} {\cal M}_\odot$, while the percentage decreased to 1-3\% for groups of mass larger than $10^{14} \ h^{-1} {\cal M}_\odot$, the  latter result 
having been confirmed using the photometric Sloan Digital Sky Survey Data Release 2 (SDSS DR2). 
Numerical simulations
by \cite{Donghia} predicted a higher fraction of fossil systems (33\%) amongst groups of mass $\sim 10^{14}  \ h^{-1} {\cal M}_\odot$. \cite{vonbenda07}, also using numerical simulations,
found that 24\% of groups with masses in the range $(1-5) 
\times 10^{13} \ h^{-1}{\cal M}_\odot$ were fossil groups. 
From semi-analytical models, \cite{sales07} 
estimated that fossil groups represent $(8-10)\%$ of groups with masses 
$(10^{13}-10^{15}) \ h^{-1}{\cal M}_\odot$, while \cite{Dariush07} stated that $\sim 13  \%$ of groups 
in that mass range were fossil systems and predicted that 
this percentage decreased to 3-4 \% for X-ray rich systems.

A natural question is whether the large magnitude difference between the first and second ranked galaxies
($\Delta$M$_{12} > $ 2 in the R-band), characteristic of these groups,
implies that they are a distinct class of objects or if they instead represent a tail of
the cluster (hereafter non-fossil group) distribution. To
investigate this question, \cite{Donghia} used high-resolution
N-body/hydrodynamical simulations to compare four 
simulated fossil and eight non-fossil groups, all of virial masses close
to $1 \times 10^{14} \ h^{-1} \ {\cal M}_\odot$. 
They found that the values of the
 magnitude gap, $\Delta$M$_{12}$,
for the 12 systems, were correlated with the halo assembly time, 
such that fossil groups assembled earlier than non-fossils. Similarly,
\cite{Dariush07} concluded,
by the study of fossil groups in the Millennium Simulation, 
that fossils assemble a higher fraction of their masses at higher
redshifts than non-fossil groups. 
The most accepted scenario for fossil groups is then that they are not a 
distinct class but, are instead,
examples of groups/clusters that collapsed early.

Although fossil groups in general do not appear differ from galaxy
clusters of similar masses, except for their earlier times of formation,
it was realised, observationally, that the first-ranked galaxies in
fossil and non-fossil groups differ in some respects.
First, their shapes differ: while fossil, first-ranked, galaxies are
often disky, brightest cluster galaxies are often boxy \citep{KPJ06}. 
Second, their stellar
populations are dissimilar: 
fossil group first-ranked galaxies are not as
old as brightest cluster galaxies  \citep{delaRosa08}. 
These findings motivated us to revisit
the study of \cite{Dariush07} of fossil groups in the Millennium Simulation,
but now focusing on the properties of the first-ranked galaxies (as
opposed to the complete system). 
By searching in the Millennium Simulation,  we created 
two samples, one of fossil groups 
with $\Delta$M$_{12} > 2$ (in the R-band) 
and a second control sample of non-fossil groups  with systems with
$\Delta$M$_{12} < 0.5$.  Both the simulations themselves 
and a mock catalogue were used to complete these searches.
In addition, we searched for fossil groups in the Sloan Digital Sky Survey Data Release 6 (SDSS DR6) \citep{AMSDSS08}
using the same criteria, to compare with the mock catalogue as 
a test of the semi-analytic model and also to examine observationally the results
of our searching algorithm.

The layout of this paper is as follows. 
In Sect.~\ref{construction} we briefly describe the galaxies in the Millennium
Simulation, and the search for simulated fossil and non-fossil groups. 
Section \ref{differences} contains a comparison between fossil and non-fossil groups,
and simulated and observed fossils. In particular, we
discuss the implications of
these results for the evolution of first-ranked galaxies in fossil groups. 
In Sect.~\ref{samples} we describe the construction of a mock catalogue and 
the procedures of group identification. 
We also   
perform an identification of fossil groups in SDSS DR6 and compare the results with
those obtained from the mock catalogue.
Finally, we summarise the paper in Sect.~\ref{conclusions}.

\section{Construction of the Simulated Fossil Group Sample}
\label{construction}
\subsection{Dark Matter Particles and Galaxies}

The Millennium Simulation is the largest completed cosmological
Tree-Particle-Mesh (TPM, \cite{Xu95}) $N$-body simulation
\citep{Springel+05}, which evolved 10 billion ($2160^3$) dark
matter particles of mass $8.6 \times 10^8 \ h^{-1} {\cal M}_\odot$ within
a periodic box of $500 \, h^{-1}\,\rm Mpc$ on a side, using a
comoving, softening length of $5 \, h^{-1} \, \rm kpc$.\footnote{The
Millennium Simulation, developed by the Virgo Consortium, is available at
http://www.mpa-garching.mpg.de/millennium} The cosmological parameters
of this simulation correspond to a standard cosmological model
($\Lambda$CDM): $\Omega_m=0.25$, $\Omega_\Lambda=0.75$, $\sigma_8=0.9$, 
and $h=0.73$.  
The merging history trees were stored for 60 output times separated by time intervals given by
$ ln(1+z_n)= n (n+35)/4200$. These are the basic
inputs required by the semi-analytic model.

We use a run of \cite{deLucia07}'s semi-analytic model to extract
galaxies with positions, velocities, as well as absolute magnitudes (in
five photometric bands, BVRIK) and stellar masses, among other quantities.
In this model, the branches of the halo merger tree are followed forward
in time, and several astrophysical processes are included such as gas infall
and cooling, reionization of the Universe, star formation, black hole
growth, AGN and supernova feedback, galaxy mergers, and spectro-photometric
evolution.  The final output at $z=0$ produced $\sim 10\times 10^6$
galaxies with absolute magnitudes $M_R-5 \ log \ h <-17.4$ and stellar masses larger 
than $3 \times 10^8 \ h^{-1} {\cal M}_\odot$. This version
of the semi-analytic model provides an improved fit
to the bright end of the galaxy luminosity function, compared with former model
developed by \cite{Croton+06}. This sample is called Millennium Simulation 
Galaxy Catalogue (MSGC hereafter).

The cosmological parameter set used in the Millennium Simulation 
was that inferred from the first-year WMAP (Wilkinson Microwave Anisotropy Probe Observations, \citealp{Spergel03}). 
Differences in the cosmological parameter sets 
corresponding to the $1^{st}$ and $3^{rd}$-year WMAP \citep{Spergel07} 
(mainly in $\sigma_8$, $\Omega_m$, $n$) 
produced a significant delay in structure formation in the WMAP3 case,
and the number of halos drawn for the WMAP1 cosmology to be higher.   
However, \cite{Wang08} investigated the implications of this delay for the 
observed properties of galaxies. They compared results obtained from a simulation
for the cosmological parameters of WMAP1 plus a semi-analytic model (such as MSGC) 
with those for a simulation for the cosmological parameters of WMAP3 plus 
a semi-analytic model. They found that the luminosity functions, correlation functions, 
and Tully-Fisher relations were almost identical in both cases. They stated that
the galaxy clustering and other observable properties were far more sensitive 
to the galaxy formation physics than to the cosmological parameters; the semi-analytic 
parameters then may be able to compensate for the delay in structure formation producing 
galaxy populations at z=0 that agree with observations. \cite{Wang08} 
also concluded that substantial differences between the models appear at redshifts higher than 2, which are beyond the scope of this paper.

\subsection{FoF galaxy halos}
\label{halos}

Since we were interested in studying fossil groups in the MSGC,
we identified galaxy halos by using a standard
method. Groups of galaxies in the MSGC were identified by using
a Friends-of-Friends (FoF) algorithm in real space \citep{DEFW85} with
a linking length of 0.2 of the mean particle density, which corresponds
to an overdensity of 200. We checked that all galaxies given in a galaxy halo belonged 
to the same DM halo.

Finally, only groups of galaxies with more than $10$ members were selected
We refer to these groups hereafter as FoF galaxy groups
(or halos).

For all FoF galaxy groups, we compute the velocity dispersion, virial radii, and virial theorem masses. The value of the virial theorem mass was computed to be:
\begin{eqnarray}
{\cal M}_{\rm FOF} &=& { \pi \over G}\,R_h\,\sigma_{3D}^2 
\label{mviroverl} 
\end{eqnarray}
where 
$R_h = \left \langle 1/R_{ij} \right\rangle^{-1}$ is the harmonic mean
projected separation, given the projected separations 
$R_{ij}$ (see eq. [10--23] of \citealp{BT87}). 
The virial radius is $r_{vir}=\pi R_h$ and the 3-D velocity dispersions were
calculated in the MSGC by using the peculiar velocities 


\subsection{ The Sample of  Fossil Groups} 
\label{fossils}

\begin{table}
\begin{center}
\caption{Median Properties of Fossil Groups identified in the Millennium Simulation
\label{medians}
}
\begin{tabular}{cccc}
\hline
\hline
 Property & MSGC & Mock Catalogue & SDSS DR6 \\
\hline
\\
 \#  & $729 $ & $22$ & $6$ \\
\\
${\cal M}_{FOF} \ [h^{-1}{\cal M}_\odot]$ & $7.5\times 10^{13}$ & $6.7\times 10^{13}$ & $8\times 10^{13}$  \\
\\
$\sigma_{3D} \ [ km/s ]$ & $600$ & $ 596$ & $565$   \\
\\
$r_{vir} \ [ Mpc \ h^{-1} ]$ & $0.98$ & $1.0$  & $1.1$  \\
\\
$M_R -5log(h)$ & $-23.48$ & $-23.73$ & $-22.17 *$  \\
\\
$M_* \  [h^{-1} {\cal M}_\odot] $ & $ 2.8\times 10^{11}$ & $4.2\times 10^{11}$& $-$  \\
\\
\hline
\end{tabular} 
\parbox{8cm}{
Notes: \# number of fossil groups, ${\cal M}_{FOF}$: virial mass of the FoF galaxy group, $\sigma_{3D}$: velocity dispersion of the FoF galaxy group, $r_{vir}$ : virial radius of the FoF galaxy group, $M_R - 5 log h$: rest frame R-band absolute magnitude of the brightest galaxy within $0.5 r_{vir}$ of the FoF galaxy group, $M_* $: stellar mass of the brightest galaxy.\\
$*$ In this case it is in the r-band. k+e corrections are computed from \cite{Blanton+03_AJ}
}
\end{center} 
\end{table}

To represent fossil groups, we selected FoF galaxy groups with masses larger
than $5 \times 10^{13} \ h^{-1} {\cal M}_\odot $, which contained galaxies 
with a magnitude distribution displaying a gap $\Delta$M$_{12} > $ 2 (in 
the R-band), 
when considering objects within a radius of
$0.5 r_{vir}$.
The adopted lower limit to the group mass ($5\times 10^{13} \ h^{-1} {\cal M}_\odot$)
ensured that the fossil groups chosen
were also X-Ray fossils, according to the
work of \cite{Dariush07}.

In the MSGC, $729$ FoF galaxy halos satisfy the fossil criteria,
which represent $5.5\%$
of the FoF galaxy groups in the studied mass range.
The median properties of these groups are quoted in Table~\ref{medians}.

\subsection{The Sample of Non-Fossil Groups}
The principal aim of this paper is to compare 
the properties of fossil and non-fossil 
groups. 
\cite{Dariush07} analysed the differences between fossil and 
non-fossil systems by concentrating on the global properties of the groups.
In this work, we intend to go further to not only confirm their findings but 
also extend their analyses to the brightest galaxies in these systems.
We constructed
a control sample of non-fossil groups, 
using the similar criteria to that used for fossil groups in the MSGC but
with the one difference that the magnitude gap between 
the brightest and second 
brightest galaxies within $0.5r _{vir}$ was less 
than $0.5$ (for the fossil groups this had to be larger
than 2 magnitudes). 
The control sample comprises $3786$ FoF galaxy halos. 

\begin{figure}
 \centering
 \includegraphics[width=8cm]{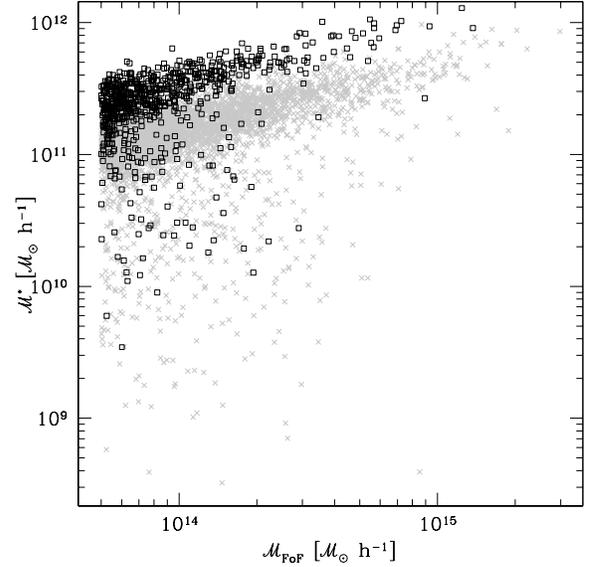}
      \caption{Scatter plot of stellar mass of the brightest galaxy and virial mass of the group. Squares correspond to 
the sample of fossil groups, while crosses correspond to non-fossil groups.}
         \label{masas}
\end{figure}

Figure \ref{masas} shows a plot of stellar masses 
versus FoF masses for both fossil (squares) and control (crosses) samples. 
To avoid inferring results dependent on the mass of the brightest galaxies 
\citep{deLucia06} from the control and the fossil samples, we selected
subsamples with approximately equal numbers of fossil and non-fossil
groups, where \emph{the distribution of stellar masses of the central galaxies
had been matched}.  This exercise provided two samples of fossil and non-fossil
groups each with about 680 objects.
We note that by insisting that the distributions of \emph{stellar masses of the 
central galaxies}  were identical for the two samples, the virial masses of the groups themselves 
were, on average, smaller for fossil than for non-fossil groups. This effect can be seen in Fig.~\ref{masas} 
if we match the y-axis distributions.
However, this procedure is justified since we were interested in
the properties of the first-ranked galaxies only. 

\section{The brightest group galaxies}
\label{differences}

\begin{figure*}
 \centering
 \includegraphics[width=12cm]{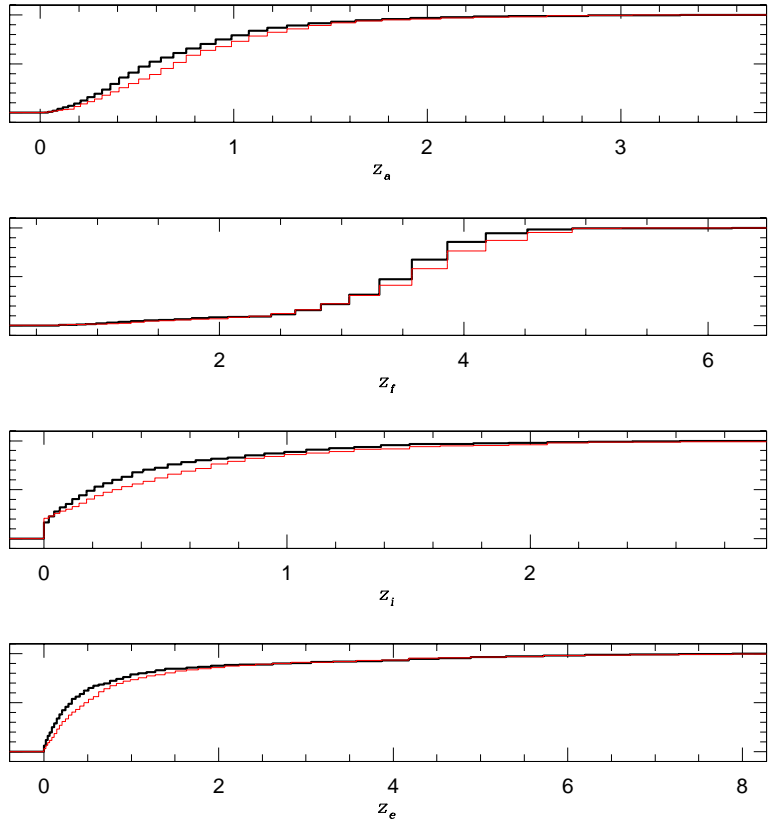}
      \caption{Cumulative distribution of assembly, formation, identity, and
extended identity times (from top to bottom). Thick lines correspond
to fossil groups, while thin lines are for non-fossil groups.}
         \label{times}
\end{figure*}

\subsection{The formation and evolution}
Given the availability of merger trees of galaxies in the Millennium
Simulation, it is possible to study the formation and evolution of
central galaxies in groups.  
\cite{deLucia06} and \cite{deLucia07}
studied the evolution of galaxies of \emph{different} stellar mass.  They defined
a set of particular times related to formation and evolution of
galaxies.  Following their work, we analyse the properties of the central
galaxies of fossil and non-fossil groups, both samples having the \emph{same} stellar
mass distributions. Briefly, the different times are defined as follows:

\begin{itemize}
\item Assembly time ($z_a$) is the time when $50\%$ of the final stellar mass was already 
contained in a single galaxy.
\item Formation time ($z_f$) is the time when half of the mass of the stars contained in
the final galaxy at redshift zero have already formed. 
\item Identity time ($z_i$) is the time when the latest major merger occurred (
major merger is adopted as ${\cal M}_1/{\cal M}_2<3$ ( ${\cal M}_1 > {\cal M}_2$) ).
\item Extended identity time ($z_e$) is the latest time when the sum of the 
masses of all progenitors merging at that time was greater than a third of the 
mass of the main progenitor (multiple simultaneous minor mergers).
\end{itemize}

\begin{table}
\begin{center}
\caption{Median redshifts and probability values from K-S test: fossil versus non-fossils
\label{KSprob}
}
\begin{tabular}{cccc}
\hline
\hline
 & Fossil & Non-Fossil & K-S\\
\hline
$z_a$ & $0.5642$ & $ 0.6871$ & $6\times 10^{-6}$ \\
$z_f$ & $3.5759$ & $3.5759$ & $6\times 10^{-6}$ \\
$z_e$ & $0.3197$ & $0.4566$ & $5\times 10^{-6}$\\
$z_i$ & $0.2798$ & $0.4566$ & $1\times 10^{-7}$\\ 
$z_a (FoF)$ & $0.4566$ & $0.1749$ & $0$ \\
\hline
\end{tabular}
\end{center} 
\end{table}
We compute all of these times for the brightest galaxies in fossil and non-fossil groups, and also the assembly time of the FoF halos. 
In Table \ref{KSprob}, we show the median of the different characteristic times and also
include the probabilities that both distributions, for fossil and non-fossil groups,
are drawn from the same distribution (Kolmogorov-Smirnov test (K-S test)).
\emph{K-S tests indicate significant differences in all cases}. However, given the shape of the distributions, 
in some cases the median value is insufficient to show these differences (for instance, when the median of $z_f$ in both samples is the same), therefore, in Fig.~\ref{times}, we show the cumulative distributions of times in Fig.~\ref{times}. Thick lines correspond
to fossil groups, while thin lines are for non-fossil groups. 
Brightest galaxies
of fossil groups have (a) assembled, 
(b) formed their stars, 
(c) experienced their last major merger and 
(d) experienced multiple simultaneous mergers, 
\emph {all at lower redshifts} than bright central galaxies in non-fossil groups. In addition, we confirm that the FoF halos of fossil groups assembled earlier than non-fossil groups.

To check if the fact that we are analysing fossil 
and non-fossil groups with 
different halo mass ranges could introduce any bias, 
we repeated the process described above by 
\emph{matching the distributions of group virial masses for fossil 
and non-fossil groups}. We then succeeded in reproducing all of the results described by
\cite{Dariush07}, in particular, that fossil groups in general,
assembled earlier than non-fossil groups, for a given range of group virial masses. 
We also found that central galaxies in fossil groups have assembled, and
experienced both their last major merger and multiple mergers 
all at lower redshifts than the central galaxies in non-fossil groups, 
but they have formed their stars at higher redshifts. This result resembles 
the analysis of \cite{deLucia06} since by matching the virial masses of the 
groups, the resulting sample of fossil groups have central galaxies with 
stellar masses typically larger than those of non-fossil groups.

Regardless of whether either the stellar masses of the central galaxies or the virial masses of the halos are matched, although the FoF galaxy halos of fossil systems have assembled earlier (\cite{Dariush07,Donghia}, confirmed in this work using an independent semi-analytic model), their central galaxies have, on average, assembled later and, even more importantly, first-ranked galaxies of fossil groups also continue to experience major mergers for a longer period of time than in non-fossil systems.


\subsection{The morphologies and morphological mix during the last major merger }

The criteria used to select fossil groups did not include any constraints
on the morphological type of the first-ranked galaxy. However, it is
interesting to study whether the first-ranked galaxy has any tendency to 
be an early-type galaxy. 

We must consider the shortcomings of the semi-analytic models 
when we attempt this analysis  given that,
historically, most semi-analytic models have failed to reproduce some trends,
although the new versions are fairly good at reproducing many observational results related to morphology (see \citealp{Bertone07}). 
However, it is still interesting to observe general trends, defining morphology 
as completed by \cite{deLucia06}. A
galaxy is classified as elliptical if $\Delta M < 0.4$ ($\Delta M =
M_{bulge}-M_{total}$ in the B-band), spiral if $\Delta M >1.56$ and S0
in between. 
These authors state that this morphological type determination is 
robust for galaxies with stellar masses larger than a few times $10^{9} \ h^{-1} {\cal M}_\odot$. 
As can be observed from Table~\ref{medians}, 
we are dealing with galaxies beyond this limit.
We find that $12\%$ of central galaxies of fossil groups
in the MSGC are non-elliptical ($5\%$ S  and $7\% $ S0's).
Similar percentages are found in the control sample ($3\%$ are S and $9\%$ are S0).

Regarding the luminosities of the first-ranked galaxies in fossil groups, 
we found that even when no specific selection criterion was used to select luminous,
central, group galaxies, the final sample of fossil 
groups had quite a bright median magnitude of $\sim -23.5$. 
We note that a search for fossil groups in the photometric SDSS \citep{Santos07}, 
using similar selection
criteria (but with no lower limit to the mass of the groups) produced
a sample of groups with a mean absolute magnitude of  $M_R - 5 \ log \ h = -23.74$

\cite{KPJ06} found that
central galaxies of fossil groups are
different in their isophotal shapes (which are often disky) 
compared with the central galaxies of
non-fossil systems (which often have boxy shapes). They then suggested
that the central galaxies of fossil
groups could be the result of wet mergers (gas-rich mergers) unlike the
galaxies in non-fossil groups.
It is, therefore, interesting to study
the morphologies of objects that represent the pre-merging subclumps
that produces the central galaxy, after a major merger.
We call the process a wet merger if at least one of the two subclumps is
a spiral galaxy, a  dry merger if both subclumps are elliptical galaxies
and mixed if one of them is elliptical and the other is an S0 or
both are S0s.  Our results are quoted in Table~\ref{mergers}. In contrast
 to expectations, 
we find that the vast majority of central galaxies in
fossil groups in the MSGC are not produced by wet major mergers.  
In fact, \emph{central galaxies
in both fossil and non-fossil groups appear to have undergone the same type
of mergers}, i.e. from the same morphological mix, 
and are mostly the result of gas-poor mergers.  
This analysis was repeated
for the subsample of systems that have ellipticals at their centres 
(88\% of the total sample) and
similar results were found (values are also quoted in Table~\ref{mergers}).

On the other hand, \cite{Khochfar05} investigated whether the observational 
isophotal shape distribution of elliptical galaxies could be reproduced in 
semi-analytic models by using both the morphology of galaxies that 
merge in the last major merger and their mass ratio. Following their work, 
we classified the central elliptical galaxies that have had a major merger 
into ``boxy'' or ``disky'' according to 
the following criteria:
\begin{itemize} 
\item Ellipticals that experience last major mergers between two bulge-dominated galaxies 
(${\cal M}_{\rm bulge} \ge 0.6 \ {\cal M}_{\rm tot}$) produce ``boxy'' remnants 
independently of the mass ratio
\item Last major mergers with mass ratio $1 \le M_1 / M_2 < 2$ produce ``boxy'' ellipticals. 
\item Last major mergers with mass ratio $2 \le M_1 / M_2 < 3$ produce ``disky'' ellipticals. 
\end{itemize}
Our results for the sample with elliptical galaxies at their centres 
that have had a major merger are quoted in Table~\ref{mergers}. 
This result disagrees with the predictions of \cite{KPJ06} based on 7 
elliptical galaxies in fossil groups.
\emph{We found no differences between the isophotal shapes of 
elliptical galaxies in fossil or non-fossil systems}, and 
in both the likelihood of being boxy was higher than being disky.
However, the particular semi-analytic model used in this work \citep{deLucia07} 
could produce a slightly higher fraction of boxy ellipticals than observations 
(for instance, for non-fossil groups we found that 
for a median B-absolute magnitude of $-21$, the ratio $N_{boxy}/N_{disky}$ 
is around $2.3$ while the observations 
of \cite{Bender92} and the predictions of \cite{Khochfar05} 
are close to $1.5$ (see Fig.~$3$ in \cite{Khochfar05}).

\begin{table}
\begin{center}
\caption{Morphological mix of subclumps that have merged to form the brightest central galaxy 
\label{mergers}
}
\begin{tabular}{ccc}
\hline
\hline
 & \multicolumn{2}{c}{Full Sample}\\
\cline{2-3}
& Fossils & non-Fossils \\
\hline
DRY (E+E) & $48\%$ & $39\%$ \\
Mixed (E+S0 or S0+S0) & $22\%$ & $26\%$ \\
WET (S+E or S+S0 or S+S) & $8\%$ & $9\%$  \\
No major merger & $22\%$ & $26\%$   \\
\hline
\hline
 & \multicolumn{2}{c}{ E galaxy at the centre }\\
\cline{2-3}
& Fossils & non-Fossils \\
\hline
DRY &   $57\%$ &  $45\%$ \\
Mixed &  $26\%$ & $31\%$  \\
WET &    $8$\% & $9\%$  \\
No major merger & $9\%$ & $15\%$  \\
\hline
boxy &  $76\%$ & $75\%$ \\
disky & $24\%$& $25\%$ \\
\hline
\end{tabular} 
\end{center} 
\end{table}
\section{Mock galaxy catalogue and observations}
\label{samples}
Although the principal aim of this paper is to study the brightest
galaxies of fossil groups identified in the Millennium Simulation, a
direct comparison of  the fossil groups identified in the both simulations
and observational catalogues can be used as a powerful test.
We have particular interest to compare the fraction of FoF systems that are fossil groups in both a mock catalogue constructed from the MSGC and the Sloan Digital Sky Survey.

To derive results that can be compared directly with observations,
we construct a mock catalogue in redshift space using a snapshot at $z=0$ of 
the MSGC. 
We compute observer frame galaxy apparent magnitudes from the
rest-frame absolute magnitudes provided by the semi-analytical model
and tabulated (k+e)-corrections from \cite{Poggianti97}, 
where the corrections
were calculated according to an evolutionary synthesis model that reproduces the 
integrated galaxy spectrum in the range $1000-25000 \AA$.
Distorted redshifts are computed using the
peculiar velocities of each galaxy.  The mock catalogue comprises 
$\sim 1.3\times 10^6$ galaxies with apparent magnitudes lower than $R=17.77$ within
the volume of one simulation box ($z_{max} \sim 0.17$, $\pi/2 \ sr$).

Groups of galaxies in the mock catalogues (FoF galaxy halos) were identified
by using an algorithm similar to that developed by \cite{Huchra82},
which is an adaptation of the FoF algorithm that takes account of the
distortion caused by peculiar motions (redshift space) and the apparent
magnitude cut-offs. As performed in real space, groups were identified as clusters with
an overdensity of 200 and above. Particular care was taken in
estimating the group centres.
We used a method that computed the
projected centre positions by weighting appropriately the positions by the local densities and 
luminosities \citep{Diaz05}.

The masses, virial radii, and velocity dispersions of the FoF galaxy group were
computed using Equation~\ref{mviroverl}, although the radial velocity dispersions ($\sigma_v$) 
where then calculated using 
the biweight estimator described by 
\citealp{BFG90} ($\sigma_{3D}=\sqrt{3} \sigma_v$). Only FoF galaxy halos with more 
than ten members were considered.

\subsection{Fossil Groups in a mock catalogue}
The criteria used to select fossil groups was described in Sect.~\ref{fossils}.
Besides considering only groups with masses larger
than $5\times 10^{13} \ h^{-1} {\cal M}_\odot$,  
we also restricted the depth of the FoF galaxy halos by selecting groups with 
redshifts lower than $0.1$. This restriction was based on statistical analyses 
that revealed that the reliability of the identification algorithm in redshift space 
increases for groups at redshifts below $z_{lim}=0.1$. 

We found a comparatively low number of fossil
groups:  $22$ FoF galaxy halos were classified as fossils, which represented
$\sim 3\%$ of the FoF galaxy groups in the mock catalogue. 
Median properties of these groups were 
also quoted in Table~\ref{medians}. For the mock sample, we estimated the mean 
density of fossil groups up to the median redshift of the sample, 
which is a fairly robust measure of density. We measured $1.4\times 10 ^{-6} h^3 \ Mpc^{-3}$.

It can be observed that the fraction of groups that satisfy the fossil criteria 
in the mock catalogue is lower than the  $\sim 5.5 \%$  found in MSGC. 
This discrepancy occurs for the following reason.
In real space (MSGC) the search for galaxies whose magnitudes respect the
$\Delta$M$_{12} > 2$ selection criterion  
is completed within a sphere, which cannot be replicated for
the mock catalogue given that we operate in redshift space, and by 
 then measuring projected distances.
When considering projected distances in the selection of fossil groups, bonafide
members in the outskirts of the groups may fail to satisfy the $\Delta$M$_{12}$ 
magnitude criterion, which then produces a smaller fraction of fossil
groups (a similar effect is seen when fossil groups 
are defined to be within $1r_{vir}$
instead of $0.5 r_{vir}$). This hypothesis was tested in the mock catalogue by 
reidentifying fossil groups in the mock catalogue, but, instead of using redshift 
space information, we used the available real space information to search for galaxies 
within a sphere of radius $0.5 r_{vir}$. 
The percentages that we found in this simple test were far more similar to those drawn from MSGC.

We note that  in observational catalogues the entire identification procedure
will be affected in a similar way as that of our mock catalogue. We must therefore recall 
that percentages or number densities
drawn from observations will be underestimated. In particular, in our study
case, we derived 3\% from the mock catalogue (which should be
equivalent to the observations) but the real number should be closer to 5.5\%
for the fraction of groups that are fossils, from the entire population
of groups/clusters with masses larger than $5 \times 10^{13} \ h^{-1}
{\cal M}_\odot$.

\subsection{Comparison with SDSS}
\label{SDSS}

 \begin{figure*} \centering \includegraphics[width=12cm]{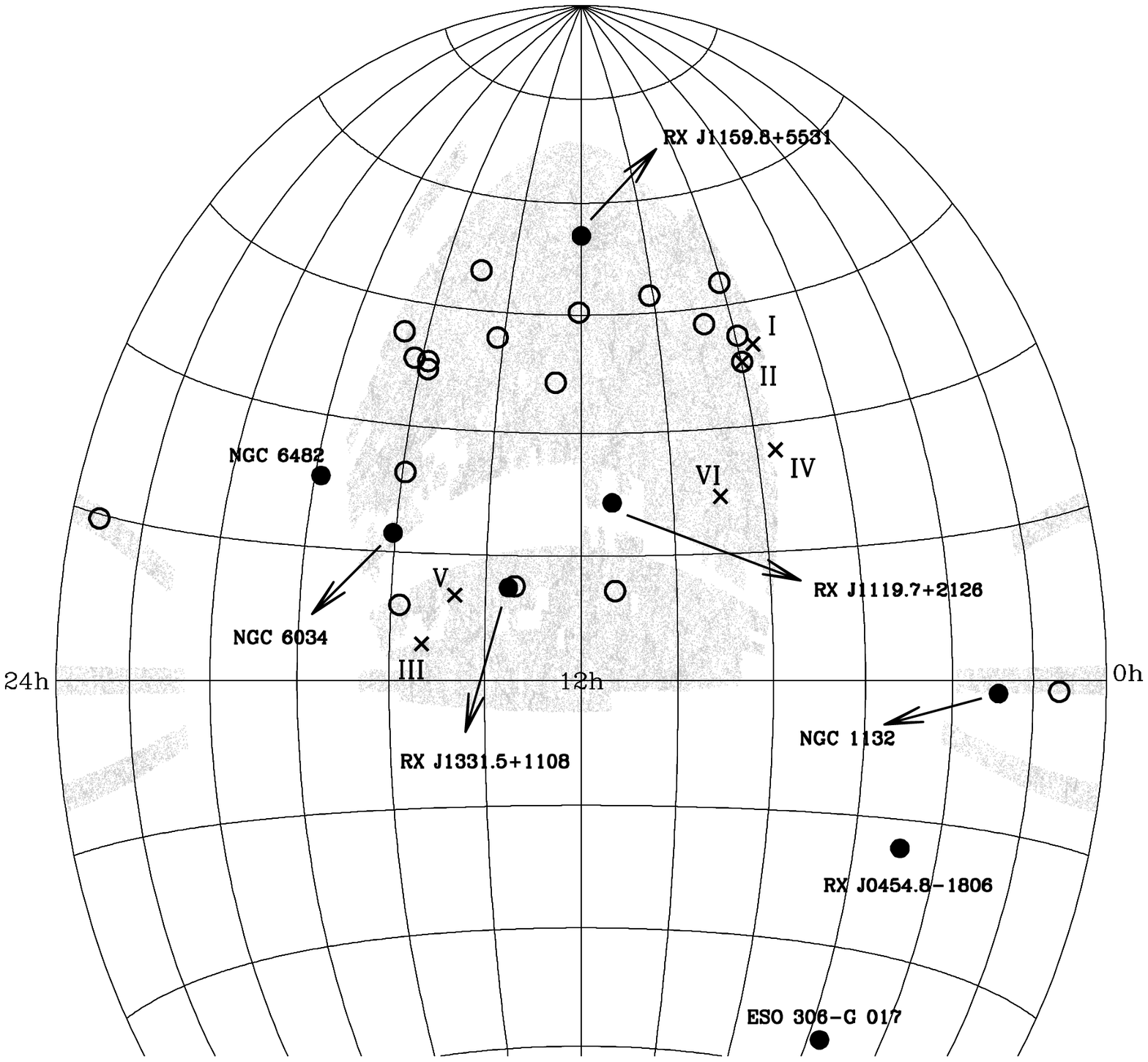}
      \caption{ Grey points are the galaxies of the main sample of
      the SDSS.
Crosses are the six fossil groups presented in this work. Filled circles
are the fossil groups found in the \citealp{Mendes06} with $z<0.1$, and
empty circles are the fossil groups identified by \citealp{Santos07}
with $z<0.1$
              }
         \label{aitoff}
   \end{figure*}

We used the main galaxy sample with spectroscopic redshifts of the SDSS DR6 
to identify observational fossil groups by using the same methodology
applied in our mock catalogue. This sample consisted of $574701$ galaxies with apparent 
magnitudes in the r-band lower than $17.77$ and a median redshift of $0.1$.
Briefly, FoF galaxy halos were identified by using an algorithm similar to 
that of \cite{Huchra82} (see \citealp{Merchan05} for further details of 
the procedure) with an overdensity contrast of 200. 
Then, halos with more than 10 members, masses larger than 
$5\times 10^{13} \ h^{-1} {\cal M}_\odot $ and redshifts lower than 0.1 were selected. 
Within this sample, we searched for fossil groups, i.e. groups with a 
magnitude gap of larger than 2 magnitudes in the r-band
 when considering galaxies within $0.5 r_{vir}$.  
We found 6 fossil groups in this sample, which represented a fraction of
$0.55\%$ of all groups 
with more than 10 members and masses larger than 
$5 \times 10^{13} \ h^{-1} {\cal M}_\odot$. 
Median properties of these fossil groups are quoted in Table~\ref{medians}. 

At first sight, this result is surprising, given that we predicted in the previous section 
that this fraction should be close to 3\%. 
To understand this low percentage of 
fossil groups in the SDSS, we carefully considered the causes of incompleteness in 
the SDSS and we then reproduce these incompleteness in our mock catalogue. 
Two problems were quickly identified: 
1) the incompleteness caused by fibre collisions and 2) the incompleteness 
caused by the fibre magnitude limits and the image deblending software.  
The former meant that a fraction of 
$\sim 70\%$ of galaxies that have a neighbour closer than $55''$ were missing.
The latter caused the spectroscopic sample to become noticeably incomplete 
for galaxies brighter than $r=14.5$.
We then performed four different mock catalogues considering 
different combinations of these incompleteness,
and we report in Table~\ref{mocks} the percentages of groups considered to be 
fossils.
It can be seen that the fibre collision effect introduces a small
bias in the percentages of fossil groups identified (which can be
noted by comparing columns ``with 
close pairs'' and ``without close pairs'', in  Table~\ref{mocks}) but, in contrast,
the lack of galaxies brighter than 14.5 
strongly bias the results, causing the fractions to
be far lower than the expected 3-4\% predicted from the full mock catalogue.
Taking the incompleteness into account, the mean density of mock fossil 
groups up to the median redshift of the sample is 
$1.2 \times 10^{-7} h^3 \ Mpc^{-3}$

To  check this result, we repeated our search for groups in the
SDSS, completed above, but now considering only galaxies in the 
magnitude range $14.5<r<17.77$. We found exactly what we expected from the
simulations:  $0.27\%$ of the FoF galaxy halos were fossil systems 
(this number can be 
directly compared with the final line in the final column of Table \ref{mocks}).
For the remaining fossil groups, 
the mean density up to the median redshift was $1.5 \times 10^{-7} h^3 \ Mpc^{-3}$, 
which was quite similar to that obtained in the mock catalogue after all biases were 
considered.

The lack of bright galaxies also caused the
few fossil groups that we identified in the SDSS DR6 to have 
first-ranked galaxies that were not as bright as those found in the semi-analytical model.

\begin{table}
\begin{center}
\caption{Percentage of Groups that are fossils in four different mock catalogues
\label{mocks}
}
\begin{tabular}{ccc}
\hline
\hline
  & with close pairs  & without close pairs\\
\hline
 $R<17.77$  & $3\% $ & $4\%$  \\
\\
\hline
$14.5<R<17.77$ & $0.2\%$ & $0.26\%$  \\
\\
\hline
\end{tabular} 
\parbox{8cm}{
Notes: Mock with close pairs: described in sect \ref{samples}. 
Mock without close pairs: reproducing the fibre collision effect. 
Both were performed with different magnitude cut-offs.
}
\end{center} 
\end{table}

\begin{sidewaystable*}
\begin{center}
\caption{Fossil groups in the main galaxy sample of the SDSS DR6
\label{SLOAN}
}
\begin{tabular}{ccccccccccccc}
\hline
\hline
 name & $\alpha_{cm}$ & $\delta_{cm}$ & $z_{cm}$ & ${\cal M}_{vir}$ 
&  $\sigma_{3D}$& $r_{vir}$& $\alpha_1$ & $\delta_1$& $z_1$ & $M_1-5log(h)$& $eClass_1$ \\
 & [h:m:s] & [d:m:s] & & [$10 ^{13} \ h^{-1} {\cal M}_\odot $] & [$km \ s^{-1}$] & 
[$Mpc \ h^{-1}$] & [h:m:s] & [d:m:s] & & [r-band]\\
\hline
\\
I & 07:35:29.04 & 39:45:21.6 & 0.0866 & 11.3& 797 & 0.86 & 07:35:36.46 & 39:45:53.57 & 0.0874 & -22.43 & -0.190478 \\
\\
\hline
\\
II & 07:58:34.32 & 37:46:12.0 & 0.0406 & 7.5 &  563 & 1.01& 07:58:28.11 & 37:47:11.87 & 0.0408 & -22.29 & -0.151681 \\
\\
\hline
\\
III& 15:19:02.64 & 04:17:49.2 & 0.0466 & 8.1 & 579 & 1.04 & 15:19:03.52 & 04:20:01.14 & 0.0468 & -22.09  & -0.177914 \\
\\
\hline
\\
IV & 07:34:33.84 & 26:53:56.4 & 0.0796 & 7.9 & 551 & 1.12 & 07:34:22.22 & 26:51:44.93 & 0.0797 & -22.35 & -0.197101 \\
\\
\hline
\\
V & 14:38:55.68 &  10:08:16.8 & 0.0552 & 7.7 & 523 & 1.21 & 14:38:47.60 & 10:07:17.26 & 0.0553 & -21.70 & -0.158786 \\
\\
\hline
\\
VI** & 08:56:37.44 & 21:52:51.6 & 0.0823 & 9.9 & 544 &  1.43 & 08:56:51.80 & 21:49:49.44 & 0.0827 & -22.17 & -0.175508 \\
\\
\hline
\end{tabular} 
\parbox{20cm}{
Notes: $\alpha_{cm}$,$\delta_{cm}$: equatorial coordinates 
of the centre of mass of the group; $z_{cm}$: spectroscopic redshift of the centre of mass; 
${\cal M}_{vir}$: virial mass of the group; $\sigma_{3D}$: 3-D velocity dispersion 
of the group; $r_{vir}$: virial radius of the group; 
$\alpha_1$, $\delta_1$, $z_1$ equatorial coordinates and spectroscopic redshift 
of the brightest galaxy; 
$M_1-5log(h)$: rest frame absolute magnitude of the brightest galaxy. k+e corrections from \citealp{Blanton+03_AJ};
$eClass_1$: spectral type parameter of the brightest galaxy, provided by SDSS.\\
$**$ It is not a fossil group (see text) 
}
\end{center} 
\end{sidewaystable*}

Despite the small number of fossil groups identified in the main spectroscopic 
galaxy sample of SDSS, we attempted to analyse 
their main properties. 
In Table~\ref{SLOAN}, the properties of each of the six groups are quoted. 
Figure \ref{aitoff} shows the angular distribution of the equatorial coordinates 
$\alpha$ versus $\delta$ for all known fossil groups. 
Grey points are the galaxies in the main sample of the SDSS. 
Crosses are the six fossil groups presented in this work. 
Filled circles are the fossil groups listed in \cite{Mendes06} with $z<0.1$ and empty circles are the fossil groups identified by \cite{Santos07} in the photometric sample of SDSS with $z<0.1$.
There are three filled circles in the SDSS area that we failed to identify. 
Using the SDSS DR6 Finding Chart Tool \footnote{http://cas.sdss.org/dr6/en/tools/chart/chart.asp},
we analysed a field about the centre of those fossil groups and 
realised that they could not be identified because of different reasons. 
RX J1331.5+1108 could not be identified because its main galaxy could not be observed due to possible 
fibre collisions. 
We failed to identify RX J1159.8+5531 as a fossil group because 
the two-magnitude gap was not present in this group. We identified this group instead to be 
massive and containing numerous galaxies, but the two-magnitude criterion 
caused us to
reject this halo (the absolute magnitude difference in R-band 
between the first and second ranked galaxy is $\Delta M_{12}=1.6$). Finally,
NGC 6034 was not detected since we did not identify a massive and numerous halo 
in its position. Instead, we identified the central galaxy to be part of the outskirts 
of a larger group. A closer inspection also indicated that this galaxy also 
had a close bright neighbour in the same large group, 
causing the magnitude gap to 
be $\Delta M_{12}=1.07$ and not 2, 
as the strict definition of fossil groups requires.

It is also interesting to compare the fossil groups identified by
\cite{Santos07} and ourselves. Only one of six groups (number II in
our Table \ref{SLOAN}) was in their catalogue. One reason for these differences 
is that \cite{Santos07} insisted
that all fossil groups had extended X-ray emission in the ROSAT all-sky
catalogue.
For our six groups, one has extended X-ray emission (group II)
and one has point source X-ray emission (as detected by XMM, group number IV) 
and four do not have any detectable ROSAT emission (and have not been 
observed by other satellites).  Given that the sensitivity of ROSAT
is quite low, these groups should be studied again with deeper X-ray observations to confirm
their nature as fossil groups. Another important discrepancy between the
two works is the way in which a fossil-system is defined. Differences in the radii used to
test the magnitude gap criterion (virial radius or fixed radius in kpc) and the 
way of defining the group membership may cause each study to select different systems.

By using the NASA/IPAC Extragalactic Database 
\footnote{http://nedwww.ipac.caltech.edu}, we also analysed whether the 
SDSS fossil groups presented in this work  had been previously 
identified as groups or clusters of galaxies.
Fossils I, IV, V, and VI did not have previous identifications as systems
of galaxies. Fossil II corresponded to the X-ray cluster NGC 2484 GROUP
\citep{Popesso04} and fossil III was previously identified as the
cluster CAN 245 \citep{Wegner99}.
 
We also tested the possibility that our fossil groups were 
spurious identifications due to the incompleteness caused by the fibre collisions 
or the fibre magnitude limit that affect the spectroscopic sample of the SDSS.
Using the SDSS DR6 Finding Chart Tool, we analysed a field of 0.5 $r_{vir}$ around
the centre of each fossil group and found that
in fossil VI the brightest galaxy of this group has a close neighbour 
that is a spectroscopic target not observed due to the fibre collision problem. 
The magnitude gap between these two galaxies is less than two magnitudes 
($\Delta M_{12} \sim 1.33$). 
Although the redshift of the second-ranked galaxy is unknown, 
we note that it is likely that this is not a fossil group. This is one
of the groups that are not detected in X-rays. 
A similar situation applies to fossil group III, since in the area corresponding to
half the virial radius there is a galaxy whose (apparent) magnitude differs by less 
than 2 from the (apparent) magnitude of the brightest galaxy, 
but this galaxy was missing from the SDSS survey because it 
has a very close neighbour. However, in this case, the visual inspection is insufficient to
confirm whether this galaxy belongs to the same group.

In summary, the spectroscopic sample of SDSS DR6 is ,probably, not the most appropriate for  
the of study observational fossil groups due to its inherent 
incompleteness, which means that we are often unable to or falsely identify groups. The enormity of this galaxy data set is still however a useful guide for deeper optical observational studies or even X-ray observations, which are beyond the scope of this paper. 

\section{Conclusions}
\label{conclusions}
To help interpret observations, 
we have analysed the properties of the first-ranked galaxies 
of simulated fossil groups and their 
main differences with respect to the brightest galaxies 
of non-fossil systems. To perform this analysis, 
we have studied a sample of fossil groups obtained from the largest simulated galaxy 
catalogue at present, the Millennium simulation combined with a semi-analytic model 
of galaxy formation \citep{deLucia07}. 
The sample of fossil groups identified was sufficiently large to enable robust 
statistical analyses to be completed.
A control sample of non-fossil groups was also selected.
The goal of this research was twofold: 
to predict results based on semi-analytical models about fossil groups and test the semi-analytical model of formation and evolution by comparing with observational results.

Our analysis was performed in both the Millennium Galaxy Catalogue (MSGC) and a
mock catalogue. The fraction of fossils predicted by the mock catalogue was
compared with the fraction of fossils found in the SDSS data set by applying the same
algorithm of identification. If the two main sources of incompleteness 
affecting the spectroscopic sample of SDSS were taken into account,
 the fraction of fossil groups
found in the mock catalogue was consistent with that observed in the SDSS.
We have presented six candidate fossil systems in SDSS DR6. 
One has been rejected after visually analysing the surroundings of the brightest galaxy. 
The remaining groups await X-rays observations or deeper photometric studies 
to confirm their nature as fossil groups.

Our main results can be summarised as follows:
we confirm the old age feature of fossil systems and that fossil 
groups with masses larger than $5 \times 10^{13} \ h^{-1} {\cal M}_\odot$ represent 
$\sim 5.5\%$ of massive systems in the same mass range, in 
the Millennium Galaxy Catalogue and a lower 
percentage (of 3\%) of similar 
systems in a mock galaxy catalogue. Fossils in the Mock Catalogue were identified 
in redshift space (just as achieved in observations), 
where galaxies from the outskirts of the groups affect 
the $\Delta M_{12}$ selection criterion and reduce the
resulting number of fossils.
We found that $88\%$ of fossil systems have central galaxies that are ellipticals.
In addition, we found that 
the first-ranked galaxies of fossil groups and non-fossil groups 
have the same morphological mixtures,  which can be considered to be the results of
gas poor mergers. 
If the central galaxy of a fossil group is always elliptical, the likelihood that its progenitor galaxies merge in a dry merger is higher. Although this result disagrees with the observational studies of \cite{KPJ06} for seven elliptical galaxies in fossil groups, further observational data is required to fully resolve this issue.
%
%
Finally, we investigated the nature of the central galaxies by 
keeping track of their merging histories. On the one hand, the fossil groups in general 
assembled earlier than non-fossil groups, whilst their central galaxies assembled later. We also found that
first-ranked galaxies in fossil groups have undergone a major merger later than 
their counterparts in non-fossil systems. We expect this result to be confirmed by future
observational catalogues.

\begin{acknowledgements}
We thank the anonymous referee for comments that helped to improve this work.
We thank Dr. Laerte Sodr\'e Jr., Walter dos Santos and Dr. Ariel Zandivarez 
for many useful discussions in all phases of this project. 
This work was partially supported by the European Commission's ALFA-II programme
through its funding of the Latin-American European Network for
Astrophysics and Cosmology (LENAC), Consejo de Investigaciones Cient\'{\i}ficas
y T\'ecnicas de la Rep\'ublica Argentina (CONICET) and Secretar\'{\i}a de Ciencia y 
T\'ecnica, UNC (SeCyT).
CMdO acknowledges financial help from FAPESP through the
thematic project 01/07342-7.
The Millennium Simulation databases used in this paper and the web 
application providing online access to them were constructed as part of the 
activities of the German Astrophysical Virtual Observatory.
\end{acknowledgements}

\bibliographystyle{aa} 
\bibliography{cgs} 

\end{document}